\documentclass[aps,pra,floatfix,twocolumn,showpacs,preprintnumbers,amsmath,amssymb,a4paper,superscriptaddress]{revtex4-1}
\usepackage{braket}
\usepackage{dsfont}
\usepackage{amsmath}
\usepackage{amssymb}
\usepackage{mathtools}
\usepackage{graphicx}
\usepackage{dcolumn}
\usepackage{bm}
\usepackage{multirow}
\usepackage{leftidx}
\usepackage{color}
 \usepackage[german,english]{babel}
\makeatletter
\newcommand{\vast}{\bBigg@{4}}
\newcommand{\Vast}{\bBigg@{5}}
\makeatother

\begin{document}
\title{Generalized local frame transformation theory for excited species in external fields}

 \author{P. Giannakeas}
 \email{pgiannak@purdue.edu}
 \affiliation{Department of Physics and Astronomy, Purdue University, West Lafayette, Indiana 47907, USA}
 
 \author{Chris H. Greene}
 \email{chgreene@purdue.edu}
 \affiliation{Department of Physics and Astronomy, Purdue University, West Lafayette, Indiana 47907, USA}
\affiliation{Purdue Quantum Center, Purdue University, West Lafayette, Indiana, 47907,
USA}

\author{F. Robicheaux}
\email{robichf@purdue.edu}
\affiliation{Department of Physics and Astronomy, Purdue University, West Lafayette,
Indiana 47907, USA}
\affiliation{Purdue Quantum Center, Purdue University, West Lafayette, Indiana, 47907,
USA}

 \date{\today}

\begin{abstract}

A rigorous theoretical framework is developed for a generalized local frame transformation theory (GLFT). 
The GLFT is applicable to the following systems:  to Rydberg atoms or molecules in an electric field, or to negative ions in any combination of electric and/or magnetic fields.
A first test application to the photoionization spectra of Rydberg atoms in an external electric field demonstrates dramatic improvement over the first version of the local frame transformation theory developed initially by Fano and Harmin.
This revised GLFT theory yields non-trivial corrections because it now includes the full on-shell Hilbert space without adopting the truncations in the original theory.
Comparisons of the semi-analytical GLFT Stark spectra with {\it ab initio} numerical simulations yields errors in the range of a few tens of MHz, an improvement over the original Fano-Harmin theory whose errors are 10-100 times larger.
Our analysis provides a systematic pathway to precisely describe the corresponding photoabsorption spectra that should be accurate enough to meet most modern experimental standards.
\end{abstract}

\maketitle
\section{Introduction}
Symmetries in a separable Hamiltonian system elegantly enable the quantum states to be described simply in terms of a few good quantum numbers.
Another intriguing class are Hamiltonian systems that possess approximate local symmetries, i.e. systems that exhibit different symmetries in different portions of the configuration space but not over the entire spatial domain.
To address this class of Hamiltonians, Fano \cite{fano1981stark} introduced the concept of a {\it{local frame transformation}} (LFT) which begins from solutions of the time-independent Schr\"odinger equation in the different portions of configuration space where the Hamiltonian obeys different symmetries, and then matches the sets of approximate ``good'' but incompatible quantum numbers via a frame transformation.
This theoretical advance has been used to interpret and quantitatively describe a plethora of diverse physical systems.
In particular, Fano \cite{fano1981stark} and Harmin \cite{harminpra1982} employed the LFT to describe Stark photoabsorption spectra of alkali metal atoms.\cite{harminpra1982,harminprl1982,harmin1990theory} The Stark effect of more complex systems could also be treated by the combination of the LFT with multichannel quantum defect theory (MQDT)\cite{seaton1983quantum}, such as Rydberg atoms with two valence electrons \cite{Armstrong1993prl,Armstrong1994pra,Robicheaux1999pra}, noble gases\cite{sakimoto1986jpb,fielding1992}, and even molecular hydrogen\cite{sakimoto1989jbp}.

Furthermore, the LFT theory gives a compact description of a variety of physical processes such as dielectronic recombination \cite{harmin1986precise}, negative-ion photodetachment in magnetic \cite{greene1987pra} or electric fields \cite{RauWong1988pra,WongRauGreene1988pra,SlonimGreene1991} or under generic external confinement \cite{robischwi2015}, or ultracold atomic and/or dipolar collisions in the presence of external trapping potentials \cite{GrangerBlume2004prl,Giannakeas2012pra,Zhang2013pra,giannakeas2013prl}.
In molecular applications frame transformation theory has been decisive in describing the rich rovibrational Rydberg spectra of diatomic molecules \cite{greene1985molecular}, and the dissociative recombination of $H_3^+$ \cite{kokoulineprl2003,JungenPrattPRL}.

Despite the versatile landscape of physical applications and the successes of the LFT theory, it lacks one important attribute for a comprehensive theory: there is no systematic pathway for improving the accuracy of the method.
Indeed, high precision experiments on the photoabsorption spectra $^7$Li in the presence of an electric field by Stevens {\it{et al}} \cite{stevens1996precision} showed that the Fano-Harmin LFT theory \cite{fano1981stark,harminpra1982} deviates from the experimental observations by several hundreds of MHz.
In addition, several theoretical investigations have attempted to identify the origin of these discrepancies and check the range of validity of the LFT\cite{zhao12,giannakeas2015}.
In this paper, a generalized LFT (GLFT) theory is developed based on more reliable and complete physico-mathematical grounds whose rigor allows an extension of calculations to much higher accuracy.
Due to the generic scope of the GLFT, it can be equally applied to Rydberg atoms or molecules in an electric field, or to negative ions in any combination of electric and/or magnetic fields.
Following the formal derivation of this GLFT, a first test application to the Stark effect of Rydberg atoms yields Stark photoabsorption Rydberg spectra 10-100 times more accurate than the Fano-Harmin LFT theory.

This work is organized as follows: Section \ref{sec1} focuses on the formulation of the generalized local frame transformation approach addressed for generic Hamiltonians which possess different symmetries in different parts of the configuration space.
Section \ref{sec2} develops the generalized local frame transformation theory to treat the photoionization of a Rydberg atom in a uniform external electric field and clarifies differences with the Fano-Harmin theory.
Section \ref{sec3} presents calculations that illustrate the main differences between the Fano-Harmin theory and the present generalized local frame transformation approach.
Finally, Section \ref{sec4} summarizes and concludes our analysis.

\section{Generalized local frame transformation theory} \label{sec1}
The concept of local frame transformation theory is particularly aimed at systems with a nonseparable Schr\"odinger equation, but which exhibits incompatible 
symmetries in different regions of configuration space.
This type of Hamiltonian has the following form:
\begin{equation}
 H=H_0 +V_s(\boldsymbol{r})+V_c(\boldsymbol{r}),
 \label{genham}
\end{equation}
where $H_0$ denotes an unperturbed separable Hamiltonian and $V_s(\boldsymbol{r})$, $V_c(\boldsymbol{r})$ are two potential terms.
This treatment assumes that the two potential terms exhibit different symmetries, i.e. frequently $V_s$ ($V_c$) has spherical (non-spherical) symmetry.
We further assume that the length scales associated with the two potentials are well separated, whereby the full Hamiltonian $H$ exhibits two regions in the configuration space of distinct symmetry.
In the first region the potential $V_s$ dominates, typically close to origin, where the Hamiltonian $H$ exhibits spherical symmetry.
Away from the origin, the $V_c$ potential prevails, normally in the asymptotic region where $H$ is separable in a non-spherical symmetry.

These considerations imply two separate coordinate systems associated with the short-range and long-range symmetries where different approximately separable solutions of the {\it{Lippmann-Schwinger}} equation exist at each energy $E$.
\begin{eqnarray}
  &~&\ket{\Psi_\kappa}=\ket{\psi_\kappa} + \hat{G}^{\rm{phys}}_c~\hat{V}_s\ket{\Psi_\kappa} ~{\rm{for}}~r\to \infty  \label{lseq1}\\
  &~&\ket{\Phi_\lambda}=\ket{f_\lambda} + \hat{G}^{\rm{phys}}_0~\hat{V}_s \ket{\Phi_\lambda}, ~{\rm{for}}~r\to 0
 \label{lseq2}
 \end{eqnarray}
 where the terms $\ket{\Psi_\kappa}$ and $\ket{\Phi_\lambda}$ correspond to standing wave solutions (for the full problem, and for the short range potential only, respectively). Here $\hat{G}^{\rm{phys}}_c$ and $\hat{G}^{\rm{phys}}_0$ represent the principal value Green's functions of the long range, $H_c=H-V_s$, and unperturbed, $H_0$, Hamiltonians.
 These Green's functions obey the corresponding proper asymptotic boundary conditions for their respective Hamiltonians and are thus denoted as {\it{physical}} Green's functions.
The terms $\ket{\psi_\kappa}$ and $\ket{f_\lambda}$ are the regular solutions of the corresponding homogeneous Schr\"odinger equation, i.e. when $V_s\equiv0$.
Note that $\kappa$ and $\lambda$ indicate collective quantum numbers that are associated with the symmetry which is fulfilled by the potentials $V_c$ and $V_s$, respectively.

Fano's key idea in Ref.\cite{fano1981stark} was to interrelate the {\it{energy normalized}} regular solutions $\ket{\psi_\kappa}$ and $\ket{f_\lambda}$ via an energy-dependent {\it{local frame transformation}} matrix $U$ satisfying
\begin{equation}
 \ket{\psi_\kappa}=\sum_\lambda\ket{f_\lambda}U_{\lambda\kappa}^{T}.
 \label{oldlft}
\end{equation}
This is a {\it local} relationship obeyed only at small distances for each energy $E$.

Asymptotically, where the $V_c$ potential prevails, the Lippmann-Schwinger relation in Eq.~(\ref{lseq1}) provides us with the corresponding $K$-matrix, i.e. $K_{\kappa' \kappa}=-\pi \braket{\psi_{\kappa'}|\hat{V}_s|\Psi_\kappa}$ which contains the relevant physics associated with the Hamiltonian $H$.
Using the Schwinger identity the $K$-matrix can obtain the following form:
\begin{equation}
 K_{\kappa' \kappa}=-\pi\braket{\psi_{\kappa'}|\hat{V}_s \hat{M}^{-1} \hat{V}_s|\psi_{\kappa}},
\label{kmat}
\end{equation}
where $\hat{M}=\hat{V}_s-\hat{V_s}\hat{G}_0^{{\rm{phys}}}\hat{V}_s -\hat{V_s}(\hat{G}_c^{{\rm{phys}}}-\hat{G}_0^{{\rm{phys}}})\hat{V}_s$. 
Note that by adding and subtracting $\hat{G}_0^{{\rm{phys}}}$ the last term in $\hat{M}$ is an infinity-free quantity.
However, the term $\hat{G}_0^{{\rm{phys}}}$ possesses singular behavior at short distances which can be tamed by choosing an on-shell complete set of states which obey Eq.~(\ref{lseq2}).
The matrix elements of Eq.~(\ref{kmat}) are computed by introducing the complete set of on-shell $\ket{\Phi_\lambda}$ states and employing the LFT $U$-matrix from Eq.~(\ref{oldlft}).
Following this prescription the $K$-matrix in Eq.~(\ref{kmat}) obtains the following form:

\begin{equation}
 K_{\kappa' \kappa}=-\pi\sum_{\lambda' \lambda}U_{\kappa' \lambda}\braket{f_{\lambda}|V_s|\Phi_\lambda}[M^{-1} ]_{\lambda \lambda'} \braket{\Phi_{\lambda'}|V_s|f_{\lambda'}}U_{\lambda' \kappa}^{T},
 \label{lftkmat}
\end{equation}
where the matrix elements $M_{\lambda' \lambda}$ obey the relation $M_{\lambda' \lambda}=\braket{\Phi_\lambda'|\hat{M}|\Phi_\lambda}$.
The {\it{roots}} of ${\rm{det}}(M)$ are associated with all the relevant resonant structure of the $K$-matrix.

The generalized LFT (GLFT) framework presented here differs in two ways from the LFT approach:
(i) the current formulation needs {\it{only}} to frame transform the regular solutions in contrast to the conventional LFT approach where an additional frame transformation was used to connect the irregular pieces of the scattering wave functions; and
(ii) the $K$-matrix in Eq.~(\ref{lftkmat}) contains the physical Green's functions allowing us to take into account not only the physics associated with the energetically open channels but also the relevant information arising from the energetically closed channels.
The latter processes affect the accuracy of the scattering observables since they are coupled with the open channel physics through the $V_s$ potential at short distances.
Note that the concept of closed or weakly closed channel physics is absent in Fano's LFT approach since only the channels which possess a classically allowed region close to the origin are considered.

\section{Improved Fano-Harmin theory in terms of generalized local frame transformation approach} \label{sec2}
\subsection{Hamiltonian and the Improved Fano-Harmin $K$-matrix}
In order to demonstrate the rigor of the GLFT approach the application to the Stark effect of non-hydrogenic atoms is now considered.
This physical system sparked the initial formulation of the LFT by Fano \cite{fano1981stark} and Harmin \cite{harminpra1982}.
The notation introduced below closely follows the notation of Ref.\cite{harminpra1982} in order to elucidate the differences between the GLFT and the original LFT.
Note that in the following atomic units  are used unless clearly stated otherwise.

Consider a neutral alkali Rydberg atom in an external electric field.
The motion of the outermost electron of an alkali atom in the presence of an electric field is described by the following Hamiltonian (in atomic units):
\begin{equation}
 H=T +V_{\rm{s}}(\boldsymbol{r})-\frac{1}{r}+F z,
 \label{eq1}
\end{equation}
where $T$ denotes the kinetic energy operator which fulfills the relation $T=-\frac{1}{2}\nabla^2_r$, $V_{\rm{s}}(\boldsymbol{r})$ indicates the residual potential of the atom, $F$ is the strength of the electric field in the $z$-direction.

To an excellent approximation, for a typical laboratory strength electric field, the non-separable Hamiltonian, $H$, becomes separable in two limiting regions of space.
Namely, at large distances ($r > r_0$), the combined external and Coulombic potential prevails, giving a separable Schr\"odinger equation in parabolic coordinates. 
Note that the length scale $r_0$ indicates the range of the electron-ion interaction.
Then, the corresponding total scattering wave function can be expressed in a compact form via the following Lippmann-Schwinger equation:
\begin{equation}
\ket{\Psi_{\epsilon \beta^F m}}=\ket{\psi_{\epsilon \beta^F m}}+ \hat{G}^{C-S, \rm{phys}}\hat{V}_{s}\ket{\Psi_{\epsilon \beta^F m}},
\label{eq3}
\end{equation}
where $\ket{\psi_{\epsilon \beta^F m}}$ is the energy normalized regular solution of the homogeneous Schr\"odinger equation , i.e. for $\hat{V}_{s}=0$.
Due to the parabolic symmetry, the $\braket{\boldsymbol{r}|\psi_{\epsilon \beta^F m}}$ expressed in parabolic coordinates has the simple form $\braket{\boldsymbol{r}|\psi_{\epsilon \beta^F m}}=[e^{i m \phi}/\sqrt{2 \pi}]\Xi_{\beta^F m}(\xi) \Upsilon_{\beta^F m}(\eta)$ where the $\Xi_{\beta^F m}$ are the eigenfunctions of the upfield $\xi$ coordinate and $\Upsilon_{\beta^F m}$ indicate the regular solutions in the down field $\eta$ coordinate which are energy normalized at $\eta\to \infty$.
$\beta^F$ denotes the fractional charge for which the motion of the electron in the upfield parabolic coordinate is bounded.
Note that $\beta^F\equiv \beta^F(\epsilon,F,m)$ is specified at each energy $\epsilon$, field strength $F$ and azimuthal angular momentum $m$.
$\hat{G}^{C-S,\rm{phys}}\equiv [\epsilon-H+\hat{V}_{s}]^{-1}$ is the principal value Coulomb-Stark Green's function which obeys the physical boundary conditions everywhere.

At short distances the electric field is overwhelmed by the combined Coulomb and the electron-ion screening potential (i.e. $\hat{V}_s$) which both possess spherical symmetry.
This suggests that the scattering wave function exhibits approximately spherical symmetry in this region of the configuration space since the electric field is negligible.
Therefore, taking into account this symmetry the scattering wave function can be expressed in spherical coordinates as follows:
\begin{equation}
 \Phi_{\epsilon \ell m}(\boldsymbol{r})= f_{\epsilon \ell m}(\boldsymbol{r})- \tan (\pi  \mu_\ell) g_{\epsilon \ell m} (\boldsymbol{r}),
 \label{eq2}
\end{equation}
where $f_{\epsilon \ell m}(\boldsymbol{r})= Y_{\ell,m}(\hat{\boldsymbol{r}})\bar{f}_{\epsilon \ell m}(r)$ [$g_{\epsilon \ell m}(\boldsymbol{r})= Y_{\ell,m}(\hat{\boldsymbol{r}})\bar{g}_{\epsilon \ell m}(r)$] are the energy normalized regular (irregular) Coulomb functions expressed in spherical coordinates and $Y_{\ell,m}(\hat{\boldsymbol{r}})$ corresponds to the spherical harmonics.
$\ell$ ($m$) denotes the orbital (azimuthal) angular momentum and $\epsilon$ indicates the total energy of the photoelectron.
$\mu_\ell$ is the $\ell-$th quantum defect which encapsulates the influence of the residual potential of the atom on the scattering wavefunction of the outermost electron and is weakly energy dependent.
In many cases the atomic potentials are inherently complicated, however, a numerical implementation of the quantum defect theory permits us to parameterize the short ranged core potential in terms of a phase shift, i.e. the quantum defects $\mu_\ell$.
The latter is used as an input in order to obtain the scattering observables asymptotically.
Note that Eq.(\ref{eq2}) is the solution of the following Lippmann-Schwinger equation $\ket{\Phi_{\epsilon \ell m}}=\ket{f_{\ell m}}+\hat{G}^{C,\rm{smooth}}\hat{V}_{s}\ket{\Phi_{\epsilon \ell m}}$. 
$\hat{G}^{C,\rm{smooth}}=\pi \sum_\ell \ket{f_{\epsilon \ell m}}\bra{g_{\epsilon \ell m}}$ represents the {\it{smooth}} Coulomb Green's function in spherical coordinates which is free of poles and does not obey the proper asymptotic boundary conditions for $E<0$ \cite{greene1979general}.

Following the prescription which is given in the previous section, the local frame transformation [see Eq.~(\ref{oldlft})] for the Stark problem is derived by interrelating the regular solutions $\ket{f_{\epsilon \ell m}}$ and  $\ket{\psi_{\epsilon \beta^F m}}$ at short and large distances, respectively.

\begin{equation}
 \ket{\psi_{\epsilon \beta^F m}}=\sum_\ell \ket{f_{\epsilon \ell m }} [U^T(\epsilon)]_{\ell \beta^F m},
 \label{eq5}
\end{equation}
where the local frame transformation $U$ contains the effect of the Stark barrier.
Note that Eq.~(\ref{eq5}), holds only in the Coulomb zone, i.e. at distances $r \ll F^{-1/2}$.

Then from Eq.~(\ref{lftkmat}) the $K$-matrix for the Stark effect can be obtained simply by making the following substitutions in the collective quantum numbers $\kappa=(\epsilon, \beta^F, m)$ and $\lambda=(\epsilon, \ell, m)$.
The complete set of states $\ket{\Phi_\lambda}$ is provided by Eq.~(\ref{eq2}) where $\ket{\Phi_\lambda}\equiv\ket{\Phi_{\epsilon \ell m}}$.
For the frame transformation matrix elements $U$ needed in Eq.~(\ref{lftkmat}), we now insert those in Eq.~(\ref{eq5}), namely $U^T_{\lambda \kappa}\equiv U^T_{\ell \beta^F m}$ (for details see Ref.\cite{harminpra1982}) which are diagonal in $m$.
Also, the matrix elements $M_{\lambda \lambda'}$ in Eq.~(\ref{lftkmat}) for the Stark effect are defined as $M_{\lambda \lambda'}\equiv M_{\ell \ell'}$.

Under these considerations the $K$-matrix for the Stark effect reads
\begin{eqnarray}
        (\underline{K})_{\beta^F,\beta^{\prime F}}&=&-\frac{1}{\pi}\sum_{\ell \ell'} U_{\beta^F\ell m}(\epsilon) \tan \pi (\mu_\ell) M^{-1}_{\ell \ell'}\times \cr
        &~&\times \tan (\pi \mu_{\ell'}) U^T_{\ell' \beta^{'F}m}(\epsilon),
\label{eq7}
\end{eqnarray}
where the elements  $\braket{f_{\epsilon \ell m}|\hat{V}_{s}|\Phi_{\epsilon \ell' m}}$ in Eq.~(\ref{lftkmat}) obey the following relation $\braket{f_{\epsilon \ell m}|\hat{V}_{s}|\Phi_{\epsilon \ell' m}}=-\frac{\tan (\pi \mu_\ell)}{\pi} \delta_{\ell \ell'}$.

In addition, the $M_{\ell \ell'}$ elements possess singularities which are removed by adding and subtracting the physical Coulomb Green's function $\hat{G}^{C,{\rm phys}}$.
More specifically, $M_{\ell \ell'}$ obeys the relation
\begin{eqnarray}
 M_{\ell \ell'}&=&\underbrace{\braket{\Phi_{\epsilon \ell m}|\hat{V}_{s}-\hat{V}_{s}\hat{G}^{C,\rm{phys}}\hat{V}_{s}|\Phi_{\epsilon \ell' m}}}_{I_{\ell \ell'}}\cr
 &-&\underbrace{\braket{\Phi_{\epsilon \ell m}|\hat{V}_{s}(\hat{G}^{C-S,\rm{phys}}-\hat{G}^{C,\rm{phys}})\hat{V}_{s}|\Phi_{\epsilon \ell' m}}}_{\frac{1}{\pi}\tan(\pi\mu_\ell) J_{\ell \ell'}\tan(\pi \mu_{\ell'})},
 \label{eq8}
\end{eqnarray}
where {\it{the roots}} of the determinant of the $M$ matrix describe the resonant features occurring at specific values of energy and electric field strength.
Therefore, Eq.~(\ref{eq8}) contains all the physics of rescattering effects due to the core as well as phenomena induced by the Stark barrier.
Note that that the use of the physical Coulomb Green's function is chosen here since it is uniquely defined for $E<0$ in spherical or in parabolic coordinates.

In view of the importance of Eq.~(\ref{eq8}) explicit expressions are provided on the evaluation of the terms $I_{\ell \ell'}$ and $J_{\ell \ell'}$ in the following subsection.

\subsection{Evaluating of $M_{\ell \ell'}$ matrix elements}
The first term in Eq.~(\ref{eq8}) is evaluated in spherical coordinates. 
The corresponding physical Coulomb Green's function in spherical coordinates is expressed in terms of the energy normalized $(f_{\epsilon \ell m},~g_{\epsilon \ell m}$) regular and irregular solutions respectively.
Namely, we have the relation 
\begin{eqnarray}
G^{C,~{\rm{phys}}}(\boldsymbol{r},\boldsymbol{r}')&=&\underbrace{\pi\sum_\ell f_{\epsilon \ell m}(\boldsymbol{r}_<)g_{\epsilon \ell m}(\boldsymbol{r}_>)}_{G^{C,{\rm smooth}}(\boldsymbol{r},\boldsymbol{r}')}\cr
&+&\pi\cot\pi\nu\sum_\ell f_{\epsilon \ell m}(\boldsymbol{r})f_{\epsilon \ell m}(\boldsymbol{r}'),\label{sphercoulomb}
\end{eqnarray}
where the vector $\boldsymbol{r}_>$ denotes that $r_>={\rm max}(r,r')$, the $\boldsymbol{r}_<$ refers to $r_<={\rm min}(r,r')$ and $\nu=1/\sqrt{-2\epsilon}$.
Note that the physical Coulomb Green's function at negative energies vanishes as $r_> \to \infty$ and $r_< \to 0$.

Using Eq.~(\ref{sphercoulomb}), the first term in Eq.~(\ref{eq8}) reads
\begin{eqnarray}
I_{\ell \ell'}=-\frac{1}{\pi}[\tan (\pi \mu_\ell)+\cot \pi \nu \tan^2 (\pi \mu_\ell)]\delta_{\ell \ell'}
\label{eq9}
\end{eqnarray}

The second term of Eq.~(\ref{eq8})  is evaluated  in parabolic coordinates whereas the corresponding physical Coulomb and Coulomb-Stark Green's functions are expressed in terms of regular $\ket{\hat{\psi}^0_{\epsilon \alpha m}}$ and irregular $\ket{\hat{\chi}^0_{\epsilon \alpha m}}$ solutions which lag $\pi/2$ phase {\it{with respect to the origin}} and are {\it{analytic in energy}}.
The index $\alpha$ refers to the fractional charge $\beta^F$ ($\beta$) of the Coulomb-Stark (Coulomb) Hamiltonian.

In detail, consider first the case of the physical Coulomb-Stark Green's function which possesses the following form in parabolic coordinates:
\begin{eqnarray}
 G^{C-S,{\rm phys}}(\xi,\phi,\eta;\xi',\phi',\eta')&=&\pi \sum^\infty_{\beta^F} \psi_{\epsilon \beta^F m}(\xi,\phi,\eta_<)\times \cr
 &\times&\chi_{\epsilon \beta^F m}(\xi',\phi',\eta_>),
\label{eq9b}
\end{eqnarray}
where $\eta_>={\rm max}(\eta,\eta')$ and $\eta_<={\rm min}(\eta,\eta')$.
The pair solutions $(\psi_{\epsilon \beta^F m},~\chi_{\epsilon \beta^F m})$ indicates the energy normalized regular and irregular solutions of the Coulomb-Stark Hamiltonian which obey the physical boundary condition at infinity.
Namely, asymptotically the irregular $\chi_{\epsilon \beta^F m}$ functions lag by $\pi/2$ the regular ones, i.e. $\psi_{\epsilon \beta^F m}$.
By employing the multichannel quantum defect theory in parabolic coordinates, the pair of solutions $(\psi_{\epsilon \beta^F m},~\chi_{\epsilon \beta^F m})$ can be expressed in terms of an alternative basis set according to the transformation
\begin{equation}
 \Bigg(\begin{matrix}
	  \psi_{\epsilon \beta^F m} \\[0.3em]
	  \chi_{\epsilon \beta^F m} \\[0.3em]
      \end{matrix}
\Bigg)= \Bigg(\begin{matrix}
	  \frac{\sqrt{A_{\beta^F m}}}{R_{\beta^F m}}& 0           \\[0.3em]
	  \frac{R_{\beta^F m}}{\sqrt{A_{\beta^F m}}}\bar{\mathcal{G}}_{\beta^Fm}^F &  \frac{R_{\beta^F m}}{\sqrt{A_{\beta^F m}}} \\[0.3em]
      \end{matrix}
\Bigg) \Bigg(\begin{matrix}
	  \hat{\psi}^{0}_{\epsilon \beta^F m} \\[0.3em]
	  \hat{\chi}^{0}_{\epsilon \beta^F m} \\[0.3em]
      \end{matrix}
\Bigg)
\label{eq9a}
\end{equation}
where the pair solutions $(\hat{\psi}^{0}_{\epsilon \beta^F m},~\hat{\chi}^{0}_{\epsilon \beta^F m})$
are the corresponding regular and irregular functions which are analytic in energy.
Recall that the irregular functions $\hat{\chi}^{0}_{\epsilon \beta^F m}$ are chosen to lag by $\pi/2$ the regular 
$\hat{\psi}^{0}_{\epsilon \beta^F m}$ solutions {\it with respect to the origin}.
We should remark that the Fano-Harmin theory employs the energy normalized solutions.
In Eq.(\ref{eq9a}) the quantity $\sqrt{A_{\beta^F m}}$ is given by the relation $A_{\beta^F m}=\frac{2 \Gamma[(1-\beta^F )\nu +m/2+1/2]}{\nu^m \Gamma[(1-\beta^F )\nu -m/2+1/2]}$ with $\nu=1/\sqrt{-2 \epsilon}$.
The amplitude $R_{\beta^F m}$ measures the amplitude modulation of the photoelectron wavefunction due to the Stark barrier [see Eq.~(44) in Ref.\cite{harminpra1982}].
In Eq.~(\ref{eq9a}), the quantity $\bar{\mathcal{G}}^F_{\beta^Fm}$ obeys the relation:
\begin{equation}
 \frac{\bar{\mathcal{G}}^F_{\beta^Fm}}{A_{\beta^F m}}= -\frac{\cot \gamma_{\beta^F m}}{R^2_{\beta^F m}}-\frac{2 \ln \nu-\tilde{\psi}(u^-_{\beta^F m})- \tilde{\psi}(u^+_{\beta^F m})}{2\pi}
\label{eq9c}
\end{equation}
where $u^\pm_{\beta^F m}=1/2\pm m/2+(1-\beta^F)\nu$ and $\tilde{\psi}(\cdot)$ denotes the digamma function.
The phase $\gamma_{\beta^F m}$ was introduced by Harmin {\it{et al}} \cite{harminpra1982} as a consequence of the Stark barrier effect and is the relative phase between the regular and irregular function which are energy normalized with respect to the origin.
Details concerning the calculation of the $R_{\beta^F m }$ amplitudes and the $\gamma_{\beta^F m}$ phases can be found either in Ref.\cite{harminpra1982} in terms of WKB theory or in Ref.\cite{giannakeas2015} in the framework of R-matrix theory.

Using Eq. ~(\ref{eq9a}) the physical Coulomb-Stark Green's function can be expressed in terms of the pair of solutions
 $(\hat{\psi}^{0}_{\epsilon \beta^F m},~\hat{\chi}^{0}_{\epsilon \beta^F m})$ which are analytic in energy.
\begin{eqnarray}
 &G&^{C-S,{\rm phys}}(\boldsymbol{r};\boldsymbol{r}')=\pi \overbrace{\sum^\infty_{\beta^F} \hat{\psi}^{0}_{\epsilon \beta^F m}(\xi,\phi,\eta_<)\hat{\chi}^{0}_{\epsilon \beta^F m}(\xi',\phi',\eta_>)}^{G^{C-S,{\rm smooth}}(\boldsymbol{r},\boldsymbol{r}')}\cr
 &+& \pi \sum^\infty_{\beta^F} \hat{\psi}^{0}_{\epsilon \beta^F m}(\xi,\phi,\eta)\bar{\mathcal{G}}^F_{\beta^Fm}\hat{\psi}^{0}_{\epsilon \beta^F m}(\xi',\phi',\eta'),
\label{eq9b}
\end{eqnarray}
where the first term indicates the the Coulomb-Stark smooth Green's function in parabolic coordinates.
Recall that the term smooth implies that the corresponding Green's function is free of poles and does not obey physical boundary conditions at infinity.

Following the same arguments for $F=0$ the physical Coulomb Green's function can be constructed in the same way as we showed for the Coulomb-Stark Green's function.
Namely, in parabolic coordinates the physical Coulomb Green's function, i.e. $F=0$, is expressed in terms of a pair solutions  $(\hat{\psi}^{0}_{\epsilon \beta m},~\hat{\chi}^{0}_{\epsilon \beta m})$ which are analytic in energy and with respect to the irregular solutions $\hat{\chi}^{0}_{\epsilon \beta m}$ lag by $\pi/2$ the regular ones, i.e. $\hat{\psi}^{0}_{\epsilon \beta m}$.
Under this assumption the physical Coulomb Green's function in parabolic coordinates obtains the following form:
\begin{eqnarray}
 &G&^{C,{\rm phys}}(\boldsymbol{r},\boldsymbol{r}')=\pi \overbrace{\sum^\infty_{\beta} \hat{\psi}^{0}_{\epsilon \beta m}(\xi,\phi,\eta_<)\hat{\chi}^{0}_{\epsilon \beta m}(\xi',\phi',\eta_>)}^{G^{C,{\rm smooth}}(\boldsymbol{r},\boldsymbol{r}')}\cr
 &+& \pi \sum^\infty_{\beta} \hat{\psi}^{0}_{\epsilon \beta m}(\xi,\phi,\eta)\bar{\mathcal{G}}_{\beta m}\hat{\psi}^{0}_{\epsilon \beta m}(\xi',\phi',\eta'),
\label{eq9b}
\end{eqnarray}
where the first term is the smooth Coulomb Green's function in parabolic coordinates.
Note that Eq.~(\ref{eq9b}) is the same as the Coulomb Green's function in Eq.~(\ref{sphercoulomb}), since both of them satisfy the same Schr\"odinger equation over the entire configuration space.
The quantity $\bar{\mathcal{G}}_{\beta m}$ obeys the relation:
\begin{equation}
  \frac{\bar{\mathcal{G}}_{\beta m}}{A_{\beta m}}=\cot \pi\nu -\frac{2 \ln \nu-\tilde{\psi}(u^-_{\beta m})- \tilde{\psi}(u^+_{\beta m})}{2\pi},
\label{eq9d}
\end{equation}
Where $A_{\beta m}=\frac{2 \Gamma[(1-\beta)\nu +m/2+1/2]}{\nu^m \Gamma[(1-\beta)\nu -m/2+1/2]}$ is the energy normalization constant with $\nu=1/\sqrt{-2\epsilon}$.
The terms $u^\pm$ are given by the relation $u^\pm_{\beta m}=1/2\pm m/2+(1-\beta)\nu$ with $\tilde{\psi}(\cdot)$ indicating the digamma function.
Finally, the difference of Coulomb-Stark and Coulomb Green's functions reads
\begin{eqnarray}
&~&G^{C-S,\rm{phys}}(\boldsymbol{r},\boldsymbol{r}')-G^{C,\rm{phys}}(\boldsymbol{r},\boldsymbol{r}')= \cr
&~&G^{C-S,{\rm smooth}}(\boldsymbol{r},\boldsymbol{r}')-G^{C,{\rm smooth}}(\boldsymbol{r},\boldsymbol{r}')\cr
&+& \pi \sum^\infty_{\beta^F} \hat{\psi}^{0}_{\epsilon \beta^F m}(\xi,\phi,\eta)\bar{\mathcal{G}}^F_{\beta^Fm}\hat{\psi}^{0}_{\epsilon \beta^F m}(\xi',\phi',\eta')\cr
&-& \pi \sum^\infty_{\beta} \hat{\psi}^{0}_{\epsilon \beta m}(\xi,\phi,\eta)\bar{\mathcal{G}}_{\beta m}\hat{\psi}^{0}_{\epsilon \beta m}(\xi',\phi',\eta')
 \label{eqgreen}
\end{eqnarray}

Eq.~(\ref{eqgreen}) can be used to evaluate the matrix elements $\braket{\Phi_\ell|\hat{V}_s(\hat{G}^{C-S,{\rm phys}}-\hat{G}^{C,{\rm phys}})\hat{V}_s|\Phi_{\ell'}}$. Due to the short-range potential $\hat{V}_s$, Eq.(\ref{eqgreen}) needs to be evaluated at small distances.
In this regime the first two terms of Eq.~(\ref{eqgreen}) cancel.
Indeed, by means of a Taylor expansion it can be shown that the two smooth Green's functions are equal in the lowest order since they are independent of the $\beta$ and $\beta^F$ fractional charges.
The validity of this approximation is ensured due to the length scale separation of the Hamiltonian $H$.

An explicit expression for the quantity $J_{\ell \ell'}$ in the second term of Eq.(\ref{eq8}) is obtained by employing the relations in Eqs.~(\ref{eq5}) and (\ref{eqgreen}).
Then the quantity $J_{\ell \ell'}$ reads
\begin{eqnarray}
 J_{\ell \ell'}&=&\sum^\infty_{\beta^F} U^{T,0}_{\ell \beta^F m}(\epsilon)\frac{\bar{\mathcal{G}}^F_{\beta^Fm}}{A_{\beta^F m}}U^{0}_{\beta^F \ell'm}(\epsilon)\cr
 &-&\sum^\infty_{\beta} U^{T,0}_{\ell \beta m}(\epsilon)\frac{\bar{\mathcal{G}}_{\beta m}}{A_{\beta m}}U^{0}_{\beta \ell'm}(\epsilon),
\label{eq10}
 \end{eqnarray}
where $U^0_{\alpha \ell m}(\epsilon)$ (with  $\alpha=\beta^F$ or $\beta$) denotes the short-range local frame transformation amplitudes that omit effects of the Stark barrier (see Eq.~(20) in Ref.\cite{harminpra1982}) whereas the frame transformations $U^0$ and $U$ from Eq.~(\ref{eq5}) obey the relation $U_{\alpha \ell m}(\epsilon)=U^0_{\alpha \ell m}(\epsilon)/R_{\alpha m}$.
For $\alpha=\beta$, the pure Coulomb amplitude is defined to be $R_{\beta m}\equiv 1$.

\begin{figure}[t!]
\includegraphics[width=\columnwidth]{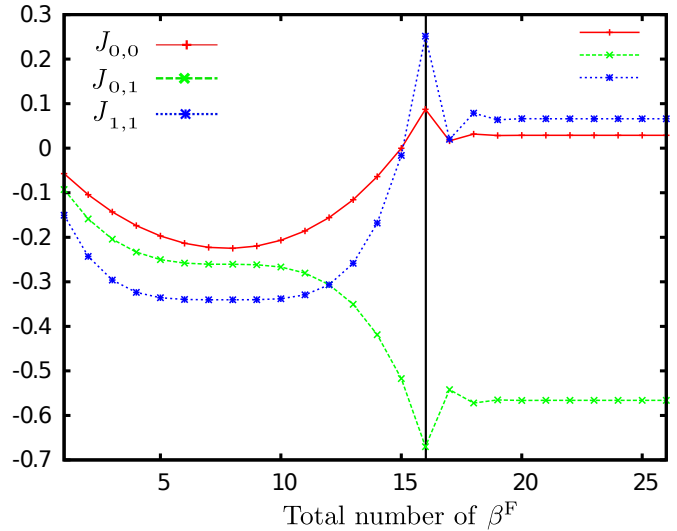}
\caption{ (color online) The quantity $J_{\ell \ell'}$ as a function of the total number of $\beta^F$ fractional charges.The field strength is $F=1000~\rm{V/cm}$, the energy is set to $\epsilon=-0.0021~\rm{a.u.}$ and the polarization of the photon is $m=0$. The three different curves correspond to different $\ell$ angular momentum combinations, i.e. $J_{0,0}$ (red solid line and crosses), $J_{0,1}$ (green dashed line and $\times$-crosses) and $J_{1,1}$ (blue dotted line and stars). The vertical black line indicates the total number of $\beta^F$ where the corresponding fractional charges $\beta^F$ are less than one. }
\label{figconv}
\end{figure}

The right hand side of Eq.~(\ref{eq10}) indicates that two infinite summations must be performed.
Actually, the difference of the two sums in Eq.~(\ref{eq10}) ensures that the left hand side remains finite.
Note that the first sum arises from the Coulomb-Stark Hamiltonian whereas the second term emerges from the Coulombic one.
For $\beta^F,~\beta\gg1$ the terms of the first sum are canceled by the terms of the second one yielding in this manner a finite $J_{\ell \ell'}$.
Intuitively, this is understood by the following: The outer classical turning point of the potential in the down field degrees of freedom $\eta$ shifts to infinity as the values of $\beta^F$ increase.
This implies that the Coulomb zone gets larger ensuring that the regular solutions vanish exponentially before tunnel out to the Coulomb-Stark region.
Recall, that the quantity $J_{\ell \ell'}$ arises only by the coherent sums over $\beta^F$ and $\beta$ of the corresponding regular functions, i.e. see the last two terms of Eq.~(\ref{eqgreen}).

Fig.\ref{figconv} illustrates the convergence of the quantity $J_{\ell \ell'}$.
More specifically, $J_{\ell \ell'}$ is plotted versus the total number of $\beta^F$ fractional charges for field strength $F=1000~\rm{V/cm}$ and energy $\epsilon=-0.0021~\rm{a.u.}$.
Note that the polarization of the photon is chosen to be parallel to the electric field, i.e. $m=0$.
The red solid line and crosses correspond to $J_{0,0}$, the green dashed line and $\times$-crosses refer to $J_{0,1}$, and the blue dotted line and stars denote the  $J_{1,1}$ matrix element.
The vertical black line corresponds to the total number of $\beta^F$ of locally ``open'' channels which is equal to 16.
This means that there are 16  different $\beta^F$ which are less than one.
One important point is that the total number of $\beta^F$ provides us with a maximum value of $\beta$, e.g. $\beta_{\rm{max}}$. 
The latter is used for the numerical convergence of the $J_{\ell \ell'}$ matrix elements according to the prescription given in the appendix \ref{converg}.

Fig.\ref{figconv} shows that the sums in $J_{\ell \ell'}$ saturate as the total number of $\beta^F$ increases beyond the total number of locally ``open'' channels, i.e. the vertical black line in Fig.\ref{figconv}.
Actually, including up to 26 locally ``open'' and ``closed'' $\beta^F$-channels, the $J_{\ell \ell'}$ is converged to six significant digits regardless the particular choice of $\ell$ angular momentum.

\begin{figure}[t!]
\includegraphics[width=\columnwidth]{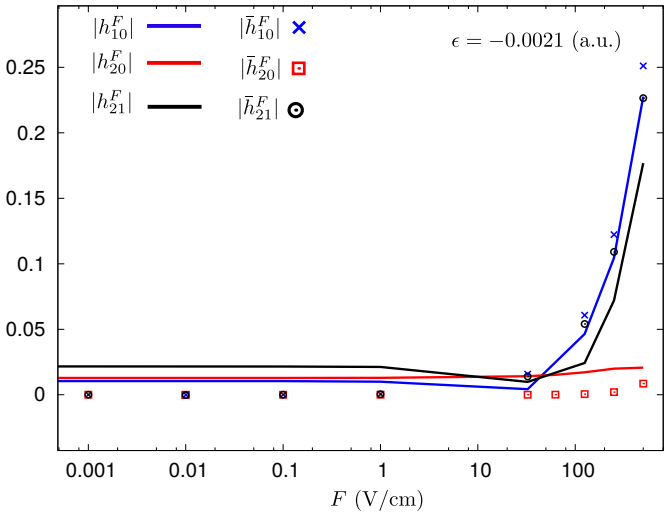}
\caption{ (color online) An illustration of the GLFT quantities $|\bar{h}^F_{\ell \ell'}|$ (scattered points) and the LFT quantities $|h^F_{\ell \ell'}|$ (solid lines) as a function of the electric field strength $F$ (${\rm V/cm}$) at an energy $\epsilon=-0.0021~({\rm a.u.})$ for  $\ell \neq \ell'$. Note that $m=0$.}
\label{fig1}
\end{figure}

\subsection{Fano-Harmin $K$-matrix and the corrections from the GLFT approach}
For reasons of completeness, a compact form of the GLFT $K$-matrix and the $K$-matrix derived from Fano-Harmin theory is introduced in this subsection.
This will permit us to unambiguously identify the main differences between the two theoretical frameworks, i.e. the LFT and GLFT.

First, we focus on the $K$-matrix in GLFT approach which is obtained by substituting in Eq.~(\ref{eq7}) the Eqs.~(\ref{eq8}),(\ref{eq9}) and (\ref{eq10}). 
After some algebraic manipulations, the $K$-matrix in the GLFT yields the following relation:
\begin{eqnarray}
 &~&K_{\beta^F,\beta^{'F}}= \sum_{\ell \ell '} 
 U_{\beta^F \ell m} [(\cot(\pi \underline{\mu})-\underline{\bar{h}}^F)^{-1}]_{\ell \ell'} U^{T}_{\ell' \beta^{'F} m},
\label{glftkmat}
\end{eqnarray}
where $\cot(\pi \underline{\mu})$ indicates a diagonal matrix whose elements fulfill the relation $\cot(\pi \underline{\mu})_{\ell \ell'}=\cot(\pi \mu_{\ell}) \delta_{\ell \ell'}$.
The matrix elements of $\underline{\bar{h}}^F$ fulfill the following relation:
\begin{eqnarray}
 \bar{h}^F_{\ell \ell'}&=&-\cot(\pi \nu) \delta_{\ell \ell'}-\sum^\infty_{\beta^F} U^{T,0}_{\ell \beta^F m}(\epsilon)\frac{\bar{\mathcal{G}}^F_{\beta^Fm}}{A_{\beta^F m}}U^{0}_{\beta^F \ell'm}(\epsilon)\cr
 &+&\sum^\infty_{\beta} U^{T,0}_{\ell \beta m}(\epsilon)\frac{\bar{\mathcal{G}}_{\beta m}}{A_{\beta m}}U^{0}_{\beta \ell'm}(\epsilon).
\label{barhfquant}
 \end{eqnarray}

Similarly, the Fano-Harmin $K$-matrix in the LFT approach has the following form
\begin{eqnarray}
 &~&K^{\rm{F-H}}_{\beta^F,\beta^{'F}}= \sum_{\ell \ell '} 
 U_{\beta^F \ell m} [(\cot(\pi \underline{\mu})-\underline{h}^F)^{-1}]_{\ell \ell'}U^{T}_{\ell' \beta^{'F} m}
 \label{hfkmat}
\end{eqnarray}
where the matrix elements $h_{\ell \ell'}^F$ is a coherent sum over the physical $\beta$-channels. Namely,
\begin{equation}
h^F_{\ell \ell'}=\sum_{\beta^F}^{\beta^F<1} U^{T,0}_{\ell \beta^F m}(\epsilon) \frac{\cot \gamma_{\beta^F m}}{R^2_{\beta^F m}}U^{0}_{\beta^F \ell' m}(\epsilon).
\label{harmin}
\end{equation}

Comparing the $K$ matrices from GLFT and Fano-Harmin approach, i.e. Eqs.~(\ref{glftkmat}) and (\ref{hfkmat}) respectively we observe that both possess the same functional form whereas the quantities $\bar{h}^F_{\ell \ell'}$ (see Eq.~\ref{barhfquant}) and $h^F_{\ell \ell'}$ (see Eq.~(\ref{harmin}) encapsulate all the relevant information for the resonant features of the photoabsorption spectra.
However, the quantities $\bar{h}^F_{\ell \ell'}$ and $h^F_{\ell \ell'}$ encompass the main differences between the two approaches.
More specifically, we observe that additional terms emerge in the improved Fano-Harmin theory (i.e. GLFT approach).
The additional terms arise from two important classes of corrections: (I) Fano-Harmin theory includes only the $\beta$-channels which possess a well in the Coulomb dominated zone, i.e. the locally {\it{open}} channels.
Indeed, in Eq.~(\ref{harmin}) the quantity $h^F_{\ell \ell'}$ the coherent sum over the $\beta$-channels is taken up-to $\beta$ values less than one.
This indicates that in $h^F_{\ell \ell'}$ only the locally open channels are taken into consideration.
On the contrary, in the present theory all the channels are included and it is indicated by the fact the coherent sum in Eq.~(\ref{barhfquant}) goes to infinity. 
(II) Fano-Harmin theory assumes that the {\it{smooth}} Coulomb Green's function expressed in spherical coordinates is approximately equal to the {\it{smooth}} Coulomb Green's function expressed in parabolic coordinates which in turn is equated to the {\it{smooth}} Coulomb-Stark Green's function in parabolic coordinates (for a detailed discussion see Ref.\cite{giannakeas2015}).
On the contrary, here this assumption is dropped yielding additional corrections as they are shown in the first term of Eq.~(\ref{barhfquant}).
The corrections arise from using the identity that only the physical Coulomb Green's function is the same in spherical and parabolic coordinates and not smooth ones as Fano-Harmin theory suggests.

\begin{figure*}[t!]
\centering
 \includegraphics[scale=0.9]{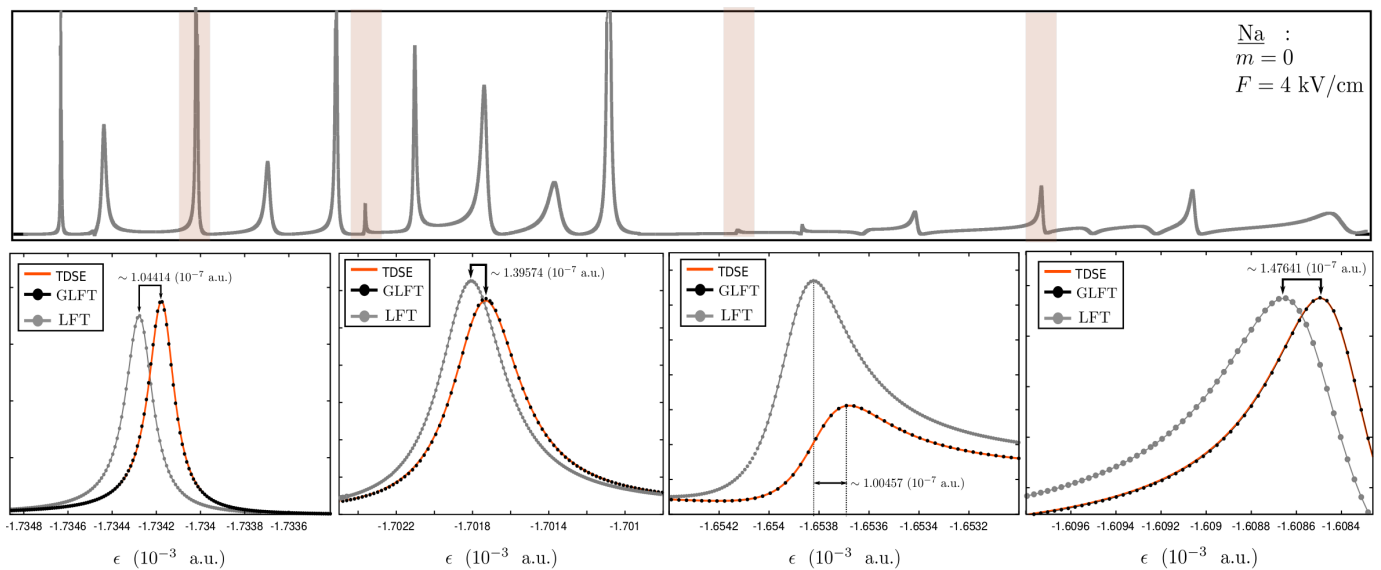}
 \caption{(Color online) The upper panel illustrates the single pulse photoionization cross section for Na atoms versus energy $\epsilon$ for $F=4~{\rm{kV/cm}}$ where the photon's polarization is parallel to the electric field.
 The shaded areas from left to right refer to resonances depicted in the lower panel where three different methods are compared. The orange solid line (TDSE) denotes the full numerical calculations whereas the gray dots refer to the Fano-Harmin theory (LFT) results and the black dots indicate the GLFT.
 Note that the arrow brackets denote the absolute difference in resonance energies between the LFT and GLFT.
 In addition, in the lower panel all the calculations are scaled by the same factor such that the far right resonance peak are the same in LFT, GLFT and TDSE approach. }
\label{fig2}
\end{figure*}

\section{results and discussion} \label{sec3}

In order to most simply demonstrate the improvements of the GLFT method over the Fano-Harmin theory, we initially focus on the regime of vanishing electric fields.
In particular the behavior of $|\bar{h}^F_{\ell \ell'}|$ and $|h^F_{\ell \ell'}|$ is studied for the case of $m=0$.
Recall that the physical origin of these quantities is that they describe the Stark field-induced resonant features of the photoionization spectra.
For non-zero fields $|\bar{h}^F_{\ell \ell'}|$ and $|h^F_{\ell \ell'}|$ for $\ell \neq \ell'$ are non-zero since the electric field couples the different angular momenta.
Therefore, for $F\to 0$, the $|\bar{h}^F_{\ell \ell'}|$ and $|h^F_{\ell \ell'}|$ for $\ell \neq \ell'$ should vanish as well.
However, as Fig.\ref{fig1} illustrates at an energy $\epsilon=-0.0021~({\rm a.u.})$ that the quantity  $|h^F_{\ell \ell'}|$ (solid lines) saturates to a constant value for vanishing field.
This implies that deeply in the linear Stark regime the Fano-Harmin theory violates the conservation of angular momentum which stems from the fact in Fano-Harmin theory that only the physical $\beta$-channels are considered.
Note that physical $\beta$-channels refer to potential curves in the down-field degree of freedom, $\eta$ parabolic coordinate, which possess a classically allowed region at small distances.
On the other hand,  the $|\bar{h}^F_{\ell \ell'}|$ (see scattered points in Fig.\ref{fig1}) vanishes for fields $F\leq1$ (V/cm).

As a second test, we compare the results from GLFT to that of a full, numerical solution of the Schr\"odinger equation.
This permits us to investigate the level of the accuracy of the present theory in the regime where the adjacent Stark manifolds are strongly mixed.
More specifically, photoionization from a $3p$-state Na atoms is considered in the presence of a field $F=4~{\rm{kV/cm}}$ where the outermost electron is ionized by a single photon.
The polarization of the photon is chosen to be parallel to the external field, i.e. $m=0$.
In order to highlight the importance of the non-trivial corrections between the GLFT and LFT approaches, the effects of electron reduced mass and mass polarization as well as spin-orbit couplings are neglected.
In addition, the appendix \ref{corepotential} gives the details of the core potential $\hat{V}_s$ for a Na atom which is used in our analysis.

Figure \ref{fig2}(upper panel) illustrates the single photon ionization cross section (in arbitrary units) for Na atoms in a $3p$-excited state as a function of the energy $\epsilon$ (in atomic units) for $F=4$~kV/cm.
The depicted spectrum is above the classical ionization threshold in an energy regime where the adjacent Stark manifolds are strongly mixed.
This permits us to compare the Fano-Harmin theory (LFT) and that derived from the GLFT (improved Fano-Harmin theory) together with the {\it{ab initio}} numerical methods.
Note that the quantum defects $\mu_\ell$ are an input parameter in LFT and GLFT approaches and they are numerically calculated by the core potential given in appendix \ref{corepotential}.
Using the same core potential for the numerical calculations the time-dependent Schr\"odinger equation (TDSE) is solved by means of standard techniques (see Ref.\cite{TR1}) completely uncorrelated with the LFT and improved GLFT approaches. 

The upper panel of Fig.\ref{fig2} shows the photoionization spectrum over the large energy scale, i.e. $\Delta \epsilon=44.1~{\rm{cm}}^{-1}$.
In this large energy scale the LFT, GLFT the TDSE photoionization spectra are indistinguishable.
However, differences become apparent when the comparison is performed on a finer energy scale.
The shaded areas in the upper panel of Fig.\ref{fig2} correspond from left to right to zoomed-in figures in the lower panels.
Also the spectra of LFT, GLFT and TDSE are scaled by the same factor which is chosen such that all peaks of the far right resonance in the lower panel of Fig.\ref{fig2} are the same between LFT, GLFT and TDSE approaches.

In the lower panels the gray (black) dots indicate the LFT/Fano-Harmin theory (GLFT/improved Fano-Harmin theory) and the orange solid line indicates the TDSE method.
We observe that the resonant energies for the LFT and its improved version GLFT disagree by more than $500~{\rm{MHz}}$ (or $7.61035\times 10^{-8}$ a.u.) (see bracket arrows).
More specifically, from left to right in the lower panels of Fig.\ref{fig2} the absolute errors indicated by the square brackets are $686,~917,~660,~970$ MHz.
In comparison, the GLFT is in excellent agreement with the numerical TDSE results, with absolute errors in resonance positions of $5.5,~20,~5,~0.8$ MHz or in atomic units $8.37138\times 10^{-10},~3.04414\times 10^{-9},~7.61035\times 10^{-10},~1.21766\times 10^{-10}$ from left to right in the lower panel.
Clearly the GLFT improves by more than an order of magnitude over the Fano-Harmin theory for strongly mixed Stark manifolds.

Another feature which is depicted in the lowest panel of Fig.\ref{fig2} is that the LFT theory exhibits discrepancies also in the amplitude of the photoionization cross-section.
Starting from left to right, the amplitude discrepancies between LFT and TDSE in the first two resonant peaks are about $\sim 6-9 \%$.
On the other hand, the amplitude discrepancies between TDSE and GLFT are less than $\sim 1\%$.
This trend is fulfilled also in the extreme case of the third resonance whose amplitude is significantly smaller with respect to the rest of the photionization spectra (see the upper panel of Fig.\ref{fig2}).
However, the LFT calculations in this case yield a resonance peak which is twice as big as in the GLFT and/or TDSE results.

Similar tests carried out for the photoabsorption Stark spectra of Li atoms show similar trends.
Zhao {\it et al.} \cite{zhao12} point out errors in the LFT claiming that they originate in the Fano-Harmin transformation of the irregular function.
However, Ref.\cite{giannakeas2015} demonstrates that those errors are far less severe than was claimed in Ref.\cite{zhao12} but they do matter for high accuracy calculations.
The GLFT approach eliminates such errors almost entirely.

\section{conclusions} \label{sec4}
The non-perturbative framework of the generalized local frame transformation theory is developed, providing a systematic pathway to improve the accuracy of Fano's ideas.
The present development can treat a broad class of Hamiltonians which possess local symmetries in different regimes of the configuration space due to its generic derivation.
As a first test application, the GLFT approach applied to the Stark effect documents the role of correction terms which are shown to yield significantly improved accuracy over the original Fano-Harmin LFT approach. 
Incorporation of these corrections yields photoabsorption spectra 10-100 times more accurate than the Fano-Harmin theory.
The GLFT agrees with essentially exact numerical simulations to better than a few tens of MHz.
This range of precision is readily achievable in current generation experiments \cite{stevens1996precision}.
The fact that the improved Fano-Harmin theory is based on the GLFT allows us to easily include relativistic and magnetic effects such as spin-orbit and hyperfine coupling by means of simple recoupling frame transformations.
Thus, the treatment of the Stark effect of heavy alkali atoms is a straightforward extension of the calculations reported here.
Moreover, the present approach can be applied to the Stark effect of multichannel Rydberg spectra such as the alkaline earth metal atoms or molecular Rydberg states\cite{Armstrong1993prl,Armstrong1994pra,Robicheaux1999pra,sakimoto1989influence}.
Another potential application is the investigation of the Stark effect of quasi-one-dimensional Rydberg atoms; an experimentally achievable concept \cite{hillerpra}.
Also, the present theory might pave an insightful avenue towards the photoionization processes of Rydberg atoms in magnetic fields \cite{mahonyprl1991}.

\begin{acknowledgements}
This work was supported by the U.S. Department of Energy, Office
of Science, Basic Energy Sciences, under Award numbers DE-SC0010545
(for PG and CHG) and DE-SC0012193 (for FR).
Some numerical calculations were performed under NSF XSEDE Resource Allocation No. TG-PHY150003.
\end{acknowledgements}

\appendix

\section{Core potential for Na atoms} \label{corepotential}
In the following, we provide the core potential of Na atoms in atomic units where its construction is based on the experimental data given by NIST atomic database \cite{nist}.
This model potential is used in the calculations of the photoionization spectrum of Na atoms in the presence of an external field.
More specifically, this potential is used in the calculations of the TDSE method and separately for the computation of the quantum defects $\mu_\ell$ 
\begin{equation}
 V_s(r)= -\frac{\alpha}{2 r^4}f_3(r)^2-\frac{Z(r)}{r}+\frac{\ell(\ell+1)}{2r^2},
\end{equation}
where $\ell$ denotes the orbital angular momentum, $\alpha$ is set to $\alpha=0.9457$ (a.u.), the quantity $Z(r)=1+f_1(r)+r f_2(r)$.
$f_3(r)$ obeys the relation:
\begin{equation}
 f_3(r)= 1-e^{-(r/r_c)^3}
\end{equation}
where the cutoff radius $r_c$ is $r_c=0.7$ (a.u.).

The quantities $f_1(r)$ and $f_2(r)$ are given by the relations:
\begin{equation}
  f_1(r)=10 e^{-\alpha_1 r}~{\rm and}~f_2(r)=\alpha_2 e^{-\alpha_3 r},
\end{equation}
where the constants $\alpha_i$ with $i=1\ldots3$ take the values $(\alpha_1,\alpha_2,\alpha_3)=(3.8538,11.0018,3.0608)$ (a.u.).

\section{Convergence and cut-off functions for $J_{\ell \ell'}$ matrix elements} \label{converg}

As we showed in the main manuscript the matrix elements $J_{\ell \ell'}$ in Eq.~(\ref{eq10}) contain two infinite summations.
However, in the numerical evaluation of $J_{\ell \ell'}$ elements the sums are truncated at a maximum $\beta$ value, e.g. $\beta_{{\rm{max}}}$. 
Under this consideration formally Eq.~(\ref{eq10}) obtains the following form:
\begin{eqnarray}
 J_{\ell \ell'}&=&\sum^\infty_{\beta^F} U^{T,0}_{\ell \beta^F m}(\epsilon)\frac{\bar{\mathcal{G}}^F_{\beta^Fm}}{A_{\beta^F m}}U^{0}_{\beta^F \ell'm}(\epsilon)\mathcal{F}_{\rm{cut-off}}(\beta^F,\beta_{\rm{max}})\cr
 &-&\sum^\infty_{\beta} U^{T,0}_{\ell \beta m}(\epsilon)\frac{\bar{\mathcal{G}}_{\beta m}}{A_{\beta m}}U^{0}_{\beta \ell'm}(\epsilon)\mathcal{F}_{\rm{cut-off}}(\beta,\beta_{\rm{max}}),
\label{apeq1}
 \end{eqnarray}
 where $\mathcal{F}(\cdot,\cdot)$ denotes a cut-off function.
 
 The particular form of the cut-off function affects the speed of the convergence of the $J_{\ell \ell'}$ matrix elements.
 For example, the choice of a step function as cut-off function, i.e. $ \mathcal{F}_{\rm{cut-off}}(x,\beta_{\rm{max}})=\Theta(\beta_{\rm{max}}-x)$, yields a slow convergence of the $J_{\ell \ell'}$ matrix elements due to Gibbs oscillations.
 Therefore, in order to accelerate the convergence of $J_{\ell \ell'}$ the following cut-off function is employed:
 \begin{equation}
 \mathcal{F}_{\rm{cut-off}}(x,\beta_{\rm{max}})=
  \begin{cases}
    1,       & \quad  x \le 1\\
    e^{-16.1 \left(\frac{x-1}{\beta_{\rm{max}}-1}\right)^b}, & \quad  x> 1,
  \end{cases}
\end{equation}
where this particular choice serves as a {\it smooth} step function and the constant $b$ takes the values $4,~6,{\rm or}~8$ ensuring that the matrix elements in Eq.~(\ref{apeq1}) are converged up to 6 significant digits.

\bibliography{qdtbiblio}

\begin{thebibliography}{33}%
\makeatletter
\providecommand \@ifxundefined [1]{%
 \@ifx{#1\undefined}
}%
\providecommand \@ifnum [1]{%
 \ifnum #1\expandafter \@firstoftwo
 \else \expandafter \@secondoftwo
 \fi
}%
\providecommand \@ifx [1]{%
 \ifx #1\expandafter \@firstoftwo
 \else \expandafter \@secondoftwo
 \fi
}%
\providecommand \natexlab [1]{#1}%
\providecommand \enquote  [1]{``#1''}%
\providecommand \bibnamefont  [1]{#1}%
\providecommand \bibfnamefont [1]{#1}%
\providecommand \citenamefont [1]{#1}%
\providecommand \href@noop [0]{\@secondoftwo}%
\providecommand \href [0]{\begingroup \@sanitize@url \@href}%
\providecommand \@href[1]{\@@startlink{#1}\@@href}%
\providecommand \@@href[1]{\endgroup#1\@@endlink}%
\providecommand \@sanitize@url [0]{\catcode `\\12\catcode `\$12\catcode
  `\&12\catcode `\#12\catcode `\^12\catcode `\_12\catcode `\%12\relax}%
\providecommand \@@startlink[1]{}%
\providecommand \@@endlink[0]{}%
\providecommand \url  [0]{\begingroup\@sanitize@url \@url }%
\providecommand \@url [1]{\endgroup\@href {#1}{\urlprefix }}%
\providecommand \urlprefix  [0]{URL }%
\providecommand \Eprint [0]{\href }%
\providecommand \doibase [0]{http://dx.doi.org/}%
\providecommand \selectlanguage [0]{\@gobble}%
\providecommand \bibinfo  [0]{\@secondoftwo}%
\providecommand \bibfield  [0]{\@secondoftwo}%
\providecommand \translation [1]{[#1]}%
\providecommand \BibitemOpen [0]{}%
\providecommand \bibitemStop [0]{}%
\providecommand \bibitemNoStop [0]{.\EOS\space}%
\providecommand \EOS [0]{\spacefactor3000\relax}%
\providecommand \BibitemShut  [1]{\csname bibitem#1\endcsname}%
\let\auto@bib@innerbib\@empty
\bibitem [{\citenamefont {Fano}(1981)}]{fano1981stark}%
  \BibitemOpen
  \bibfield  {author} {\bibinfo {author} {\bibfnamefont {U.}~\bibnamefont
  {Fano}},\ }\href@noop {} {\bibfield  {journal} {\bibinfo  {journal} {Phys.
  Rev. A}\ }\textbf {\bibinfo {volume} {24}},\ \bibinfo {pages} {619} (\bibinfo
  {year} {1981})}\BibitemShut {NoStop}%
\bibitem [{\citenamefont {Harmin}(1982{\natexlab{a}})}]{harminpra1982}%
  \BibitemOpen
  \bibfield  {author} {\bibinfo {author} {\bibfnamefont {D.~A.}\ \bibnamefont
  {Harmin}},\ }\href {\doibase 10.1103/PhysRevA.26.2656} {\bibfield  {journal}
  {\bibinfo  {journal} {Phys. Rev. A}\ }\textbf {\bibinfo {volume} {26}},\
  \bibinfo {pages} {2656} (\bibinfo {year} {1982}{\natexlab{a}})}\BibitemShut
  {NoStop}%
\bibitem [{\citenamefont {Harmin}(1982{\natexlab{b}})}]{harminprl1982}%
  \BibitemOpen
  \bibfield  {author} {\bibinfo {author} {\bibfnamefont {D.~A.}\ \bibnamefont
  {Harmin}},\ }\href {\doibase 10.1103/PhysRevLett.49.128} {\bibfield
  {journal} {\bibinfo  {journal} {Phys. Rev. Lett.}\ }\textbf {\bibinfo
  {volume} {49}},\ \bibinfo {pages} {128} (\bibinfo {year}
  {1982}{\natexlab{b}})}\BibitemShut {NoStop}%
\bibitem [{\citenamefont {Harmin}(1990)}]{harmin1990theory}%
  \BibitemOpen
  \bibfield  {author} {\bibinfo {author} {\bibfnamefont {D.~A.}\ \bibnamefont
  {Harmin}},\ }\enquote {\bibinfo {title} {Atoms in strong fields},}\ \
  (\bibinfo  {publisher} {Springer US},\ \bibinfo {address} {Boston, MA},\
  \bibinfo {year} {1990})\ pp.\ \bibinfo {pages} {61--106}\BibitemShut
  {NoStop}%
\bibitem [{\citenamefont {Seaton}(1983)}]{seaton1983quantum}%
  \BibitemOpen
  \bibfield  {author} {\bibinfo {author} {\bibfnamefont {M.}~\bibnamefont
  {Seaton}},\ }\href@noop {} {\bibfield  {journal} {\bibinfo  {journal}
  {Reports on Progress in Physics}\ }\textbf {\bibinfo {volume} {46}},\
  \bibinfo {pages} {167} (\bibinfo {year} {1983})}\BibitemShut {NoStop}%
\bibitem [{\citenamefont {Armstrong}\ \emph {et~al.}(1993)\citenamefont
  {Armstrong}, \citenamefont {Greene}, \citenamefont {Wood},\ and\
  \citenamefont {Cooper}}]{Armstrong1993prl}%
  \BibitemOpen
  \bibfield  {author} {\bibinfo {author} {\bibfnamefont {D.~J.}\ \bibnamefont
  {Armstrong}}, \bibinfo {author} {\bibfnamefont {C.~H.}\ \bibnamefont
  {Greene}}, \bibinfo {author} {\bibfnamefont {R.~P.}\ \bibnamefont {Wood}}, \
  and\ \bibinfo {author} {\bibfnamefont {J.}~\bibnamefont {Cooper}},\ }\href
  {\doibase 10.1103/PhysRevLett.70.2379} {\bibfield  {journal} {\bibinfo
  {journal} {Phys. Rev. Lett.}\ }\textbf {\bibinfo {volume} {70}},\ \bibinfo
  {pages} {2379} (\bibinfo {year} {1993})}\BibitemShut {NoStop}%
\bibitem [{\citenamefont {Armstrong}\ and\ \citenamefont
  {Greene}(1994)}]{Armstrong1994pra}%
  \BibitemOpen
  \bibfield  {author} {\bibinfo {author} {\bibfnamefont {D.~J.}\ \bibnamefont
  {Armstrong}}\ and\ \bibinfo {author} {\bibfnamefont {C.~H.}\ \bibnamefont
  {Greene}},\ }\href {\doibase 10.1103/PhysRevA.50.4956} {\bibfield  {journal}
  {\bibinfo  {journal} {Phys. Rev. A}\ }\textbf {\bibinfo {volume} {50}},\
  \bibinfo {pages} {4956} (\bibinfo {year} {1994})}\BibitemShut {NoStop}%
\bibitem [{\citenamefont {Robicheaux}\ \emph {et~al.}(1999)\citenamefont
  {Robicheaux}, \citenamefont {Wesdorp},\ and\ \citenamefont
  {Noordam}}]{Robicheaux1999pra}%
  \BibitemOpen
  \bibfield  {author} {\bibinfo {author} {\bibfnamefont {F.}~\bibnamefont
  {Robicheaux}}, \bibinfo {author} {\bibfnamefont {C.}~\bibnamefont {Wesdorp}},
  \ and\ \bibinfo {author} {\bibfnamefont {L.~D.}\ \bibnamefont {Noordam}},\
  }\href {\doibase 10.1103/PhysRevA.60.1420} {\bibfield  {journal} {\bibinfo
  {journal} {Phys. Rev. A}\ }\textbf {\bibinfo {volume} {60}},\ \bibinfo
  {pages} {1420} (\bibinfo {year} {1999})}\BibitemShut {NoStop}%
\bibitem [{\citenamefont {Sakimoto}(1986)}]{sakimoto1986jpb}%
  \BibitemOpen
  \bibfield  {author} {\bibinfo {author} {\bibfnamefont {K.}~\bibnamefont
  {Sakimoto}},\ }\href@noop {} {\bibfield  {journal} {\bibinfo  {journal}
  {Journal of Physics B: Atomic and Molecular Physics}\ }\textbf {\bibinfo
  {volume} {19}},\ \bibinfo {pages} {3011} (\bibinfo {year}
  {1986})}\BibitemShut {NoStop}%
\bibitem [{\citenamefont {Fielding}\ and\ \citenamefont
  {Softley}(1992)}]{fielding1992}%
  \BibitemOpen
  \bibfield  {author} {\bibinfo {author} {\bibfnamefont {H.}~\bibnamefont
  {Fielding}}\ and\ \bibinfo {author} {\bibfnamefont {T.}~\bibnamefont
  {Softley}},\ }\href@noop {} {\bibfield  {journal} {\bibinfo  {journal} {J.
  Phys. B: At. Mol. Phys.}\ }\textbf {\bibinfo {volume} {25}},\ \bibinfo
  {pages} {4125} (\bibinfo {year} {1992})}\BibitemShut {NoStop}%
\bibitem [{\citenamefont {Sakimoto}(1989{\natexlab{a}})}]{sakimoto1989jbp}%
  \BibitemOpen
  \bibfield  {author} {\bibinfo {author} {\bibfnamefont {K.}~\bibnamefont
  {Sakimoto}},\ }\href {http://stacks.iop.org/0953-4075/22/i=17/a=011}
  {\bibfield  {journal} {\bibinfo  {journal} {J. Phys. B: At. Mol. Phys.}\
  }\textbf {\bibinfo {volume} {22}},\ \bibinfo {pages} {2727} (\bibinfo {year}
  {1989}{\natexlab{a}})}\BibitemShut {NoStop}%
\bibitem [{\citenamefont {Harmin}(1986)}]{harmin1986precise}%
  \BibitemOpen
  \bibfield  {author} {\bibinfo {author} {\bibfnamefont {D.~A.}\ \bibnamefont
  {Harmin}},\ }\href@noop {} {\bibfield  {journal} {\bibinfo  {journal} {Phys.
  Rev. Lett.}\ }\textbf {\bibinfo {volume} {57}},\ \bibinfo {pages} {1570}
  (\bibinfo {year} {1986})}\BibitemShut {NoStop}%
\bibitem [{\citenamefont {Greene}(1987)}]{greene1987pra}%
  \BibitemOpen
  \bibfield  {author} {\bibinfo {author} {\bibfnamefont {C.~H.}\ \bibnamefont
  {Greene}},\ }\href {\doibase 10.1103/PhysRevA.36.4236} {\bibfield  {journal}
  {\bibinfo  {journal} {Phys. Rev. A}\ }\textbf {\bibinfo {volume} {36}},\
  \bibinfo {pages} {4236} (\bibinfo {year} {1987})}\BibitemShut {NoStop}%
\bibitem [{\citenamefont {Rau}\ and\ \citenamefont
  {Wong}(1988)}]{RauWong1988pra}%
  \BibitemOpen
  \bibfield  {author} {\bibinfo {author} {\bibfnamefont {A.~R.~P.}\
  \bibnamefont {Rau}}\ and\ \bibinfo {author} {\bibfnamefont {H.~Y.}\
  \bibnamefont {Wong}},\ }\href {\doibase 10.1103/PhysRevA.37.632} {\bibfield
  {journal} {\bibinfo  {journal} {Phys. Rev. A}\ }\textbf {\bibinfo {volume}
  {37}},\ \bibinfo {pages} {632} (\bibinfo {year} {1988})}\BibitemShut
  {NoStop}%
\bibitem [{\citenamefont {Wong}\ \emph {et~al.}(1988)\citenamefont {Wong},
  \citenamefont {Rau},\ and\ \citenamefont {Greene}}]{WongRauGreene1988pra}%
  \BibitemOpen
  \bibfield  {author} {\bibinfo {author} {\bibfnamefont {H.~Y.}\ \bibnamefont
  {Wong}}, \bibinfo {author} {\bibfnamefont {A.~R.~P.}\ \bibnamefont {Rau}}, \
  and\ \bibinfo {author} {\bibfnamefont {C.~H.}\ \bibnamefont {Greene}},\
  }\href {\doibase 10.1103/PhysRevA.37.2393} {\bibfield  {journal} {\bibinfo
  {journal} {Phys. Rev. A}\ }\textbf {\bibinfo {volume} {37}},\ \bibinfo
  {pages} {2393} (\bibinfo {year} {1988})}\BibitemShut {NoStop}%
\bibitem [{\citenamefont {Slonim}\ and\ \citenamefont
  {Greene}(1991)}]{SlonimGreene1991}%
  \BibitemOpen
  \bibfield  {author} {\bibinfo {author} {\bibfnamefont {V.~Z.}\ \bibnamefont
  {Slonim}}\ and\ \bibinfo {author} {\bibfnamefont {C.~H.}\ \bibnamefont
  {Greene}},\ }\href {\doibase 10.1080/10420159108211499} {\bibfield  {journal}
  {\bibinfo  {journal} {Radiation effects and defects in solids}\ }\textbf
  {\bibinfo {volume} {122}},\ \bibinfo {pages} {679} (\bibinfo {year}
  {1991})}\BibitemShut {NoStop}%
\bibitem [{\citenamefont {Robicheaux}\ \emph {et~al.}(2015)\citenamefont
  {Robicheaux}, \citenamefont {Giannakeas},\ and\ \citenamefont
  {Greene}}]{robischwi2015}%
  \BibitemOpen
  \bibfield  {author} {\bibinfo {author} {\bibfnamefont {F.}~\bibnamefont
  {Robicheaux}}, \bibinfo {author} {\bibfnamefont {P.}~\bibnamefont
  {Giannakeas}}, \ and\ \bibinfo {author} {\bibfnamefont {C.~H.}\ \bibnamefont
  {Greene}},\ }\href {\doibase 10.1103/PhysRevA.92.022711} {\bibfield
  {journal} {\bibinfo  {journal} {Phys. Rev. A}\ }\textbf {\bibinfo {volume}
  {92}},\ \bibinfo {pages} {022711} (\bibinfo {year} {2015})}\BibitemShut
  {NoStop}%
\bibitem [{\citenamefont {Granger}\ and\ \citenamefont
  {Blume}(2004)}]{GrangerBlume2004prl}%
  \BibitemOpen
  \bibfield  {author} {\bibinfo {author} {\bibfnamefont {B.~E.}\ \bibnamefont
  {Granger}}\ and\ \bibinfo {author} {\bibfnamefont {D.}~\bibnamefont
  {Blume}},\ }\href {\doibase 10.1103/PhysRevLett.92.133202} {\bibfield
  {journal} {\bibinfo  {journal} {Phys. Rev. Lett}\ }\textbf {\bibinfo {volume}
  {92}},\ \bibinfo {pages} {133202} (\bibinfo {year} {2004})}\BibitemShut
  {NoStop}%
\bibitem [{\citenamefont {Giannakeas}\ \emph {et~al.}(2012)\citenamefont
  {Giannakeas}, \citenamefont {Diakonos},\ and\ \citenamefont
  {Schmelcher}}]{Giannakeas2012pra}%
  \BibitemOpen
  \bibfield  {author} {\bibinfo {author} {\bibfnamefont {P.}~\bibnamefont
  {Giannakeas}}, \bibinfo {author} {\bibfnamefont {F.~K.}\ \bibnamefont
  {Diakonos}}, \ and\ \bibinfo {author} {\bibfnamefont {P.}~\bibnamefont
  {Schmelcher}},\ }\href@noop {} {\bibfield  {journal} {\bibinfo  {journal}
  {Phys. Rev. A}\ }\textbf {\bibinfo {volume} {86}},\ \bibinfo {pages} {042703}
  (\bibinfo {year} {2012})}\BibitemShut {NoStop}%
\bibitem [{\citenamefont {Zhang}\ and\ \citenamefont
  {Greene}(2013)}]{Zhang2013pra}%
  \BibitemOpen
  \bibfield  {author} {\bibinfo {author} {\bibfnamefont {C.}~\bibnamefont
  {Zhang}}\ and\ \bibinfo {author} {\bibfnamefont {C.~H.}\ \bibnamefont
  {Greene}},\ }\href@noop {} {\bibfield  {journal} {\bibinfo  {journal} {Phys.
  Rev. A}\ }\textbf {\bibinfo {volume} {88}},\ \bibinfo {pages} {012715}
  (\bibinfo {year} {2013})}\BibitemShut {NoStop}%
\bibitem [{\citenamefont {Giannakeas}\ \emph {et~al.}(2013)\citenamefont
  {Giannakeas}, \citenamefont {Melezhik},\ and\ \citenamefont
  {Schmelcher}}]{giannakeas2013prl}%
  \BibitemOpen
  \bibfield  {author} {\bibinfo {author} {\bibfnamefont {P.}~\bibnamefont
  {Giannakeas}}, \bibinfo {author} {\bibfnamefont {V.~S.}\ \bibnamefont
  {Melezhik}}, \ and\ \bibinfo {author} {\bibfnamefont {P.}~\bibnamefont
  {Schmelcher}},\ }\href
  {http://link.aps.org/doi/10.1103/PhysRevLett.111.183201} {\bibfield
  {journal} {\bibinfo  {journal} {Phys. Rev. Lett.}\ }\textbf {\bibinfo
  {volume} {111}},\ \bibinfo {pages} {183201} (\bibinfo {year}
  {2013})}\BibitemShut {NoStop}%
\bibitem [{\citenamefont {Greene}\ and\ \citenamefont
  {Jungen}(1985)}]{greene1985molecular}%
  \BibitemOpen
  \bibfield  {author} {\bibinfo {author} {\bibfnamefont {C.~H.}\ \bibnamefont
  {Greene}}\ and\ \bibinfo {author} {\bibfnamefont {C.}~\bibnamefont
  {Jungen}},\ }\href@noop {} {\bibfield  {journal} {\bibinfo  {journal} {Adv.
  At. Mol. Phys.}\ }\textbf {\bibinfo {volume} {21}},\ \bibinfo {pages} {51}
  (\bibinfo {year} {1985})}\BibitemShut {NoStop}%
\bibitem [{\citenamefont {Kokoouline}\ and\ \citenamefont
  {Greene}(2003)}]{kokoulineprl2003}%
  \BibitemOpen
  \bibfield  {author} {\bibinfo {author} {\bibfnamefont {V.}~\bibnamefont
  {Kokoouline}}\ and\ \bibinfo {author} {\bibfnamefont {C.~H.}\ \bibnamefont
  {Greene}},\ }\href {\doibase 10.1103/PhysRevLett.90.133201} {\bibfield
  {journal} {\bibinfo  {journal} {Phys. Rev. Lett.}\ }\textbf {\bibinfo
  {volume} {90}},\ \bibinfo {pages} {133201} (\bibinfo {year}
  {2003})}\BibitemShut {NoStop}%
\bibitem [{\citenamefont {Jungen}\ and\ \citenamefont
  {Pratt}(2009)}]{JungenPrattPRL}%
  \BibitemOpen
  \bibfield  {author} {\bibinfo {author} {\bibfnamefont {C.}~\bibnamefont
  {Jungen}}\ and\ \bibinfo {author} {\bibfnamefont {S.~T.}\ \bibnamefont
  {Pratt}},\ }\href {\doibase 10.1103/PhysRevLett.102.023201} {\bibfield
  {journal} {\bibinfo  {journal} {Phys. Rev. Lett.}\ }\textbf {\bibinfo
  {volume} {102}},\ \bibinfo {pages} {023201} (\bibinfo {year}
  {2009})}\BibitemShut {NoStop}%
\bibitem [{\citenamefont {Stevens}\ \emph {et~al.}(1996)\citenamefont
  {Stevens}, \citenamefont {Iu}, \citenamefont {Bergeman}, \citenamefont
  {Metcalf}, \citenamefont {Seipp}, \citenamefont {Taylor},\ and\ \citenamefont
  {Delande}}]{stevens1996precision}%
  \BibitemOpen
  \bibfield  {author} {\bibinfo {author} {\bibfnamefont {G.~D.}\ \bibnamefont
  {Stevens}}, \bibinfo {author} {\bibfnamefont {C.-H.}\ \bibnamefont {Iu}},
  \bibinfo {author} {\bibfnamefont {T.}~\bibnamefont {Bergeman}}, \bibinfo
  {author} {\bibfnamefont {H.~J.}\ \bibnamefont {Metcalf}}, \bibinfo {author}
  {\bibfnamefont {I.}~\bibnamefont {Seipp}}, \bibinfo {author} {\bibfnamefont
  {K.~T.}\ \bibnamefont {Taylor}}, \ and\ \bibinfo {author} {\bibfnamefont
  {D.}~\bibnamefont {Delande}},\ }\href {\doibase 10.1103/PhysRevA.53.1349}
  {\bibfield  {journal} {\bibinfo  {journal} {Phys. Rev. A}\ }\textbf {\bibinfo
  {volume} {53}},\ \bibinfo {pages} {1349} (\bibinfo {year}
  {1996})}\BibitemShut {NoStop}%
\bibitem [{\citenamefont {Zhao}\ \emph {et~al.}(2012)\citenamefont {Zhao},
  \citenamefont {Fabrikant}, \citenamefont {Du},\ and\ \citenamefont
  {Bordas}}]{zhao12}%
  \BibitemOpen
  \bibfield  {author} {\bibinfo {author} {\bibfnamefont {L.~B.}\ \bibnamefont
  {Zhao}}, \bibinfo {author} {\bibfnamefont {I.~I.}\ \bibnamefont {Fabrikant}},
  \bibinfo {author} {\bibfnamefont {M.~L.}\ \bibnamefont {Du}}, \ and\ \bibinfo
  {author} {\bibfnamefont {C.}~\bibnamefont {Bordas}},\ }\href {\doibase
  10.1103/PhysRevA.86.053413} {\bibfield  {journal} {\bibinfo  {journal} {Phys.
  Rev. A}\ }\textbf {\bibinfo {volume} {86}},\ \bibinfo {pages} {053413}
  (\bibinfo {year} {2012})}\BibitemShut {NoStop}%
\bibitem [{\citenamefont {Giannakeas}\ \emph {et~al.}(2015)\citenamefont
  {Giannakeas}, \citenamefont {Robicheaux},\ and\ \citenamefont
  {Greene}}]{giannakeas2015}%
  \BibitemOpen
  \bibfield  {author} {\bibinfo {author} {\bibfnamefont {P.}~\bibnamefont
  {Giannakeas}}, \bibinfo {author} {\bibfnamefont {F.}~\bibnamefont
  {Robicheaux}}, \ and\ \bibinfo {author} {\bibfnamefont {C.~H.}\ \bibnamefont
  {Greene}},\ }\href@noop {} {\bibfield  {journal} {\bibinfo  {journal} {Phys.
  Rev. A}\ }\textbf {\bibinfo {volume} {91}},\ \bibinfo {pages} {043424}
  (\bibinfo {year} {2015})}\BibitemShut {NoStop}%
\bibitem [{\citenamefont {Greene}\ \emph {et~al.}(1979)\citenamefont {Greene},
  \citenamefont {Fano},\ and\ \citenamefont {Strinati}}]{greene1979general}%
  \BibitemOpen
  \bibfield  {author} {\bibinfo {author} {\bibfnamefont {C.~H.}\ \bibnamefont
  {Greene}}, \bibinfo {author} {\bibfnamefont {U.}~\bibnamefont {Fano}}, \ and\
  \bibinfo {author} {\bibfnamefont {G.}~\bibnamefont {Strinati}},\ }\href@noop
  {} {\bibfield  {journal} {\bibinfo  {journal} {Phys. Rev. A}\ }\textbf
  {\bibinfo {volume} {19}},\ \bibinfo {pages} {1485} (\bibinfo {year}
  {1979})}\BibitemShut {NoStop}%
\bibitem [{\citenamefont {Top{\c{c}}u}\ and\ \citenamefont
  {Robicheaux}(2007)}]{TR1}%
  \BibitemOpen
  \bibfield  {author} {\bibinfo {author} {\bibfnamefont {T.}~\bibnamefont
  {Top{\c{c}}u}}\ and\ \bibinfo {author} {\bibfnamefont {F.}~\bibnamefont
  {Robicheaux}},\ }\href@noop {} {\bibfield  {journal} {\bibinfo  {journal} {J.
  Phys. B: At. Mol. Phys.}\ }\textbf {\bibinfo {volume} {40}},\ \bibinfo
  {pages} {1925} (\bibinfo {year} {2007})}\BibitemShut {NoStop}%
\bibitem [{\citenamefont
  {Sakimoto}(1989{\natexlab{b}})}]{sakimoto1989influence}%
  \BibitemOpen
  \bibfield  {author} {\bibinfo {author} {\bibfnamefont {K.}~\bibnamefont
  {Sakimoto}},\ }\href@noop {} {\bibfield  {journal} {\bibinfo  {journal} {J.
  Phys. B: At. Mol. Phys.}\ }\textbf {\bibinfo {volume} {22}},\ \bibinfo
  {pages} {2727} (\bibinfo {year} {1989}{\natexlab{b}})}\BibitemShut {NoStop}%
\bibitem [{\citenamefont {Hiller}\ \emph {et~al.}(2014)\citenamefont {Hiller},
  \citenamefont {Yoshida}, \citenamefont {Burgd\"orfer}, \citenamefont {Ye},
  \citenamefont {Zhang},\ and\ \citenamefont {Dunning}}]{hillerpra}%
  \BibitemOpen
  \bibfield  {author} {\bibinfo {author} {\bibfnamefont {M.}~\bibnamefont
  {Hiller}}, \bibinfo {author} {\bibfnamefont {S.}~\bibnamefont {Yoshida}},
  \bibinfo {author} {\bibfnamefont {J.}~\bibnamefont {Burgd\"orfer}}, \bibinfo
  {author} {\bibfnamefont {S.}~\bibnamefont {Ye}}, \bibinfo {author}
  {\bibfnamefont {X.}~\bibnamefont {Zhang}}, \ and\ \bibinfo {author}
  {\bibfnamefont {F.~B.}\ \bibnamefont {Dunning}},\ }\href {\doibase
  10.1103/PhysRevA.89.023426} {\bibfield  {journal} {\bibinfo  {journal} {Phys.
  Rev. A}\ }\textbf {\bibinfo {volume} {89}},\ \bibinfo {pages} {023426}
  (\bibinfo {year} {2014})}\BibitemShut {NoStop}%
\bibitem [{\citenamefont {O'Mahony}\ and\ \citenamefont
  {Mota-Furtado}(1991)}]{mahonyprl1991}%
  \BibitemOpen
  \bibfield  {author} {\bibinfo {author} {\bibfnamefont {P.~F.}\ \bibnamefont
  {O'Mahony}}\ and\ \bibinfo {author} {\bibfnamefont {F.}~\bibnamefont
  {Mota-Furtado}},\ }\href {\doibase 10.1103/PhysRevLett.67.2283} {\bibfield
  {journal} {\bibinfo  {journal} {Phys. Rev. Lett.}\ }\textbf {\bibinfo
  {volume} {67}},\ \bibinfo {pages} {2283} (\bibinfo {year}
  {1991})}\BibitemShut {NoStop}%
\bibitem [{\citenamefont {{Kramida~A.}}\ \emph {et~al.}(2015)\citenamefont
  {{Kramida~A.}}, \citenamefont {{Ralchenko~Yu.}}, \citenamefont {{Reader
  J.}},\ and\ \citenamefont {{NIST ASD Team}}}]{nist}%
  \BibitemOpen
  \bibfield  {author} {\bibinfo {author} {\bibnamefont {{Kramida~A.}}},
  \bibinfo {author} {\bibnamefont {{Ralchenko~Yu.}}}, \bibinfo {author}
  {\bibnamefont {{Reader J.}}}, \ and\ \bibinfo {author} {\bibnamefont {{NIST
  ASD Team}}},\ }\href {http://physics.nist.gov/asd} {\enquote {\bibinfo
  {title} {Nist atomic spectra database (version 5.3)},}\ } (\bibinfo {year}
  {2015})\BibitemShut {NoStop}%
\end{thebibliography}%

\end{document}